\def\beg{\begin{equation}}
\def\eeq{\end{equation}}
\documentstyle[12pt]{article}
\begin{document}
\begin{center}
{\Large{\bf Comments on ``Theoretical search for the nested quantum 
Hall effect of composite fermions" by Mandal and Jain, 
Phys. Rev. B 66,155302 (2002);cond-mat/0210181.}}
\vskip0.35cm
{\bf Keshav N. Shrivastava}
\vskip0.25cm
{\it School of Physics, University of Hyderabad,\\
Hyderabad  500046, India}
\end{center}

We find that a large number of parameters are used to create the 
correct fraction. The parameters are $\nu$, 1-$\nu$, $\nu^*$, 
${\bar n}$, $n$, $p$ and ${\bar p}$. Therefore, the predicted 
fractions need not be having the correct origin. The wave function describes a composite fermion which has 2p (even number) of flux
quanta attached to one electron. We find that it requires 
``decomposite" fermion which is the electron in an orbit from which 
the magnetic field has been detached. This kind of detachment (attachment) of flux quanta from (to) the electron is not 
consistent with the electromagnetic theory of light and violates 
Biot and Savart's law, as well as theory of relativity. If flux 
quanta are to be attached to the electron, we should solve the 
bound-state equation and determine the binding energy but bound-state has not been solved. The wave function given is not a solution 
of the bound-state equation. Therefore, Mandal and Jain's composite fermion (CF) model is incorrect.
\vfill
Corresponding author: keshav@mailaps.org\\
Fax: +91-40-3010145.Phone: 3010811.
\newpage
\baselineskip22pt
\noindent {\bf 1.~ Introduction}

     The quantum Hall effect is dominated by the observation of 
plateaus in the transverse resistivity, $\rho_{xy}$=h/$\nu$e$^2$ at integer values of $\nu$. In fact $\nu$ need not be an integer. The fractional values of $\nu$ such as 1/3 also occur. The claim of the composite fermion (CF) model$^1$ is that even number of flux quanta are attached to the electron which gives rise to the plateaus in the 
quantum Hall effect. The CF model proposes several series of 
fractions,
\beg
{n\over 2pn\pm1}, 1-{n\over 2pn\pm1}, {n\over 4pn\pm1}, 1-{n\over 4pn\pm1}, ... etc.
\eeq
In place of 2p, 4p, 6p, 8p, etc. ``even number" is proposed without 
any upper limit. In fact many of these fractions are in good 
agreement with the experimental data. Thus CF model is considered 
to be a strong candidate for the explanation of the quantum Hall 
effect. It is possible to make use of this ``even number" in the Laughlin's wave function so that one can obtain the fraction as well 
as the wave function of the quasiparticle which gives the quantum 
Hall effect. The expression (1) is neither a solution of the 
bound-state equation nor that of the Schr\"odinger equation.

     In this paper, we wish to point out that the assertion that 
``even number of flux quanta are attached to the electron" is 
incorrect. Hence any agreement with the experimental data is because 
of the fact that too much force has been put to get the experimental values of the fractional charge.

\noindent{\bf 2.~~The fractional charge}

{\it (a) Abundance.}

     The resistivity is given by,$\rho$=h/$\nu$e$^2$ so that $\nu$ 
gives the effective charge,
\beg
 e_{eff}=\nu e.
\eeq
When we take all of the prescribed values of the fraction, we find 
that a lot of values of (1) are the same as in experimental data. However, the series (1) give many more fractions than are observed 
in the data. At large values of the Landau level quantum number, the values given by (1) are far more than observed. Therefore, the 
abundance of fractions obtained by (1) is not in agreement with the data.

{\it(b) Parameters.}

     The charge is calculated using the parameters,
\beg
\nu, 1-\nu, \nu^*, {\bar n}, n, p, {\bar p}, etc.
\eeq
Therefore, there are far too many parameters to find the experimentally measured fractions of charges.

{\it(c) Particle-hole symmetry.}

     In the paper of Mandal and Jain$^1$,
\beg
\nu={n\over 2pn\pm1}.
\eeq
In addition,
\beg
\nu = 1 - {n\over 2pn\pm1}
\eeq
has been introduced. If the first value is 1/3, then the second value is 2/3. However, if the wave function for 1/3 is antisymmetric, then it will be symmetric for 2/3. Therefore, 1/3 will be a fermion and 2/3 will be a boson. In the CF model, there are always even number of flux quanta attached to an electron and hence quasiparticles are fermions. Therefore
particle-hole symmetry does not occur in the CF model. It is clear that there is internal inconsistency in the CF model. In the case of Laughlin's wave function 1/3 is produced and next value is 1/5 but not 2/3 because 1/3 and 1/5 are antisymmetric and 2/3 is symmetric. There is no way for Laughlin to have an even numerator with odd denominator.

{\it(d) Flux quanta attachment.}

   The even number of flux quanta must be attached to the electron 
to obtain the effective field inside the sample. Therefore, the 
field is,
\beg
B^*=B-2np\phi_o
\eeq
where $n$ is the number of electrons per unit area, $p$ is an 
integer and $\phi_o$=$hc/e$ is the unit flux. In order to attach 
even number of flux quanta, let us detach the flux from the 
electron. The electron going in a circular orbit produces a field 
as shown in Fig.1. Let us see whether it is allowed to detach the 
field from the electron. This process produces an electron current 
and a field. What is left after detaching B is called decomposite fermion (DF). The object (DF) will not be having the magnetic field 
but it has a charge. Hence Maxwell equations will not be obeyed.
The DF will not emit light nor will it absorb. Hence it will be completely invisible. Such objects are not known to exist in the 
real world. They surely do not exist in GaAs. Let us try to attach 
two flux quanta to the electron. One electron gives one field B, 
two electrons give $2B$ as shown in Fig.2. The $2B$ can not be given 
by one $e$, otherwise we are left with a DF. Such a DF without field 
is an unphysical object.

     When we put B$^*$ =0, $B$ is found to be quantized and when we 
put $B$=0, then B$^*$ is quantized. That seems to be quite alright. However, in superconductors, field outside the sample is continuous 
and inside the sample, in the mixed state, is quantized. Therefore, 
it is not clear why both fields $B$ and B$^*$ get quantized. The 
density of the CF is the same as that of the electrons or there is 
only one density in the CF model. The CF are much larger objects 
than the electrons. Therefore, their density can not be the same as 
that of the electrons$^2$. The density of the CF is therefore not internally consistent$^3$. The energy calculation has been performed 
for N=12, 16, 20, 24, etc. but these values are too small to give 
any realistic results.

    In the CF model, the fractional charge of 1/3 was produced by inventing the series $n/(2pn\pm1)$. Thus using three parameters, 
$p$, $n$ and $\pm1$, to produce 1/3 introduces too much 
arbitrariness and hence $n/(2pn\pm1)$ is not a model at all. The experimentalists$^4$ have been misguided$^5$ to identify the experimental data with the CF. In fact the experimental data is 
not related to CF.

\noindent{\bf3. ~~ Electromagnetic theory}

     A charged particle emits radiation which is composed of 
electric and magnetic field vectors, transverse to the direction 
of motion of the energy. The Maxwell equations determine the 
dynamics of the electric $\vec E$ and magnetic $\vec H$ vectors 
with charge density as a constant. Thus $\vec E$ and $\vec H$ 
are strongly coupled. The formation of a DF decouples $\vec H$
from the charge and hence from $\vec E$. At the present time 
the $\vec E$ and $\vec H$ are not decoupled. Similarly, extra 
$\vec H$ can not be attached to the charge. Therefore, CF 
formation is not permitted in the usual Maxwell equations. 
However, new type of Maxwell equations may be written but 
classical form of electromagnetic theory is in agreement with 
the experiments, and hence there is no need of new type of Maxwell
equations. The formation of CF is therefore not expected to 
agree with the present day experiments. The use of the same 
density for CF as for the electron makes the CF internally 
inconsistent. Therefore, at this time there is no hope for the CF 
model.

\noindent{\bf4.~~ Conclusions}.

     The attachment of two flux quanta to one electron is not 
correct. It violates the classical electrodynamics. The use of 
only one density for the quantization of $B$ as well as B$^*$ is internally inconsistent. Therefore, the field expression $B^*$=B-2np$\phi_o$, much needed to produce experimentally observed 
fractional charge, is not internally consistent. The same number 
of small and large objects can not fill the same area. The even 
number of flux quanta produce the Laughlin type wave functions 
but 2/3 can not come, so there is inconsistency. The classical electromagnetic theory is not uniquely produced$^6$. It is clear 
that the CF model lacks in good theoretical foundation$^7$ and 
its various elements are inconsistent$^8$. The field expression 
which attaches flux is not correct$^9$.

     Since Laughlin's wave function is limited to antisymmetry 
of the wave functions, which leads to odd number in the density, 
it may be thought that this wave function is not a theory of the
quantum Hall effect. The CF model is {\it not 
free of troubles} with the fundamentals. Therefore, the readers 
may be interested in a calculation which explains the 
experimental data without difficulty with fundamentals. Thus, 
a theory is given in a recent book$^{10}$. About 200 references 
of recent prl articles have been examined to check the 
applicability of the theory$^{11}$ and in all cases, the given 
theory agrees with the data. It may be mentioned that the 
particle-hole symmetry$^{12}$ around $\nu$=1/2 has been observed 
by Willett et al$^{13}$ but the interpretation given in ref.[13]
is not satisfactory.
\vskip1.25cm

\noindent{\bf5.~~References}
\begin{enumerate}
\item S. S. Mandal and J.K. Jain, Phys. Rev. B{\bf66}, 155302(2002);cond-mat/0210181
\item K. N. Shrivastava, cond-mat/0209666.
\item K. N. Shrivastava, cond-mat/0209057
\item I. V. Kukushkin et al, Nature {\bf415}, 409 (2002).
\item K. N. Shrivastava, cond-mat/0202459.
\item K. N. Shrivastava, cond-mat/0210238.
\item K. N. Shrivastava, cond-mat/0204627.
\item M. I. Dyakonov, Talk at NATO ARW ``Recent trends in theory 
of physical phenomena in high magnetic fields", cond-mat/0209206.
\item B. Farid, cond-mat/0003064.
\item K.N. Shrivastava, Introduction to quantum Hall effect,\\ 
      Nova Science Pub. Inc., N. Y. (2002).
\item K. N. Shrivastava, cond-mat/0201232.
\item K. N. Shrivastava, Mod. Phys. Lett. B {\bf 13}, 1087(1999).
\item R. L. Willett, K. W. West and L. N. Pfeiffer, Phys. Rev. Lett.
{\bf83}, 2624 (1999).
\end{enumerate}
\vskip0.1cm
Note: Ref.10 is available from:\\
 Nova Science Publishers, Inc.,\\
400 Oser Avenue, Suite 1600,\\
 Hauppauge, N. Y.. 11788-3619,\\
Tel.(631)-231-7269, Fax: (631)-231-8175,\\
 ISBN 1-59033-419-1 US$\$69$.\\
E-mail: novascience@Earthlink.net

\vskip5.5cm

Fig.1: The electron in a circular orbit produces a field $B$. If 
this field is detached from the electron current we are left with 
two objects, one is a charge called DF and the other is a field $B$.

Fig.2: One electron current produces a field of $B$ and another 
electron also produces another field $B$. Now detach the field of 
one electron so that DF is left and attach it to another electron 
so that CF is formed.
\end{document}